\documentclass[unsortedadress,preprint
 ,secnumarabic%
,amssymb, amsmath,nobibnotes,prd]{revtex4}
\begin{document}
\title{A Noncommutative Deformation of Topological Field Theory\footnote{This work is dedicated to
Professor
Alberto Garc\'{\i}a on the occasion of his 60th birthday.}}

\author{Hugo Garc\'{\i}a-Compe\'an
\email{\tt compean@fis.cinvestav.mx} and Pablo Paniagua}
\affiliation{Departamento de F\'{\i}sica,
Centro de Investigaci\'on y de Estudios Avanzados del IPN\\
P.O. Box 14-740, 07000 M\'exico D.F., M\'exico}

\date{\today}

\begin{abstract} 
Cohomological Yang-Mills theory is formulated on a noncommutative differentiable four
manifold through the $\theta$-deformation of its corresponding BRST algebra. The
resulting noncommutative field theory is a natural setting to define the
$\theta$-deformation of Donaldson invariants and they are interpreted as a mapping
between the Chevalley-Eilenberg homology of noncommutative spacetime and the
Chevalley-Eilenberg cohomology of noncommutative moduli of instantons. In the process we
find that in the weak coupling limit the quantum theory is localized at the moduli space
of noncommutative instantons.
\end{abstract}

\vskip -1truecm
\maketitle

\vskip -1.3truecm
\newpage

\setcounter{equation}{0}

\section{Introduction}

Quantum field theories on noncommutative spaces are very interesting theories that have recently
motivated a great deal of work (for a recent review see, \cite{dn}). Quantum field theories of
topological nature describing the cohomology of the moduli space of Yang-Mills instantons (with
certain gauge group and instanton number $k$) constitute a class of interesting theories whose
correlation functions of BRST-invariant field functionals ${\cal O}$ give rise to topological
invariants known as Donaldson polynomial invariants \cite{topoym}. A gravitational analog of
Donaldson theory has been, by the first time, constructed in Ref. \cite{wtg}. These kind of
theories are known generically as Cohomological Field Theories and it is natural to ask about
the formulation of a possible noncommutative deformation of these theories. A proposal for a
noncommutative deformation of cohomological field theory was given in Ref. \cite{nccft}, where
some relations to noncommutative solitons was found. However, in the present paper, we follow a
different approach making emphasis in the definition of topological invariants. Formulating a
noncommutative topological theory is another way of finding noncommutative topological
invariants.  Noncommutative classical invariants (as the Euler number or signature) are known in
the literature from some years ago in the context of the standard formulation of spectral
triples from noncommutative geometry \cite{connesbook} and recently has been also pursued by
using the method of Moyal product and the Seiberg-Witten map \cite{topologico}.

Noncommutative gauge theories are gauge theories on noncommutative spaces which can be
understood in terms of algebras of operators on certain Hilbert space. The
Weyl-Wigner-Moyal correspondence establishes an isomorphic relation between it and the
algebra of functions on ${\bf R}^{n}$, but instead of having the usual commutative product
of functions we have a noncommutative and associative star product $\star$ called the
Moyal product.  For instance a noncommutative deformation of space ${\bf R}^n$ where the
usual coordinates satisfy the commutation relations: $[ x^{\alpha},x^{\beta} ]
=i\theta^{\alpha \beta}$ with $x^{\alpha} \in End(L^2({\bf R}^{n}))$, is given by a
noncommutative and associative algebra ${\cal A}_{\star} \equiv {\bf R}^{n}_{\star}$ with
the usual Moyal bracket of functions.

Usually gauge fields are defined as connections on $G$ gauge bundles $E$ over spacetime manifold
$X$. In noncommutative geometry one substitutes vector bundles $E$ by the noncommutative Moyal
deformation of the right (or left) ${\cal A}$-module denoted by ${\cal E}$.
For instance if $S$ is a submanifold of $X$ and denoting ${\cal A}(S)$ as the set of sections of
the gauge bundle $E$ over $S$. ${\cal A}(S)$ can be regarded as an ${\cal A}_{\star}(X)$-module
that we denote ${\cal E}$, where ${\cal A}_{\star}(X)$ denotes the Moyal algebra of functions on
$X$. Alternatively the algebra ${\cal A}_{\star}(X)$ will be denoted simply as $X_{\theta}$.  
Thus in noncommutative geometry the gauge bundle $E$ can be replaced by ${\cal E}$. A
noncommutative gauge field $A^I_{\alpha}$ can be regarded as a connection in the
$X_{\theta}$-module ${\cal E}$.

In the present paper we formulate a noncommutative deformation of Witten's cohomological field
theory by replacing its underlying Lie algebra by its enveloping algebra following \cite{wess}.
We will show that this procedure leads to define a cohomology theory of the moduli space of
noncommutative Yang-Mills instantons \cite{ns}. However a rather different approach proposed in
Ref. \cite{schwarz} will be more useful for our purposes. Noncommutative instantons are
instantons on a noncommutative deformation of euclidean spacetime and in \cite{ns} a
noncommutative of ADHM description of instantons was proposed. This description gives rise to a
definition of the moduli space of noncommutative instantons as a smooth manifold free from small
instanton singularities, because these singularities are blow up by a smooth deformation of
ADHM equations. In Ref. \cite{sw} the smooth structure of the moduli space of noncommutative
instantons was corroborated and reinterpreted in terms of D-brane processes. The description of
the ADHM constructions and its relation to noncommutative twistor transform was given in Ref.  
\cite{anton}.

This paper is organized as follows: in section 2 we propose a noncommutative deformation of the
BRST invariant gauge theory. We will mainly follow the notation of Ref. \cite{topoym}.
Observables of this deformed theory are discussed in section 3. Section 4 is devoted to provide a
deformation of the Donaldson invariants and therefore to define noncommutative topological
invariants and to give an expression for the noncommutative Donaldson invariants in terms of an
integration of differential forms on the moduli space of noncommutative instantons. Through out
all the paper we have assumed that the metric field on the manifold $X$ is not a dynamical field
(spectator field) and it is not noncommutatively coupled to noncommutative matter fields. Finally
in section 5 we give our concluding remarks.

\section{The Noncommutative Lagrangian}

We start from a noncommutative deformation of the BRST-like invariant Lagrangian of the
usual topological Yang-Mills theory given by

$$
L = \int_X d^4 \sqrt{g} {\rm Tr} \bigg[ {1 \over 8} (F_{\alpha \beta} - \widetilde{F}_{\alpha
\beta})* (F_{\alpha \beta} - \widetilde{F}_{\alpha \beta}) +
D_{\alpha} \lambda * D^{\alpha} \phi - D_{\alpha} \eta *  \psi^{\alpha} + \lambda *
[\psi_{\alpha}\stackrel{\ast}{,} \psi^{\alpha}]
$$
\begin{equation}
 - \chi^{\alpha \beta} * \big(D_{\alpha} \psi_{\beta} - D_{\beta} \psi_{\alpha} -
\varepsilon_{\alpha \beta \gamma \delta} D^{\gamma} \psi^{\delta} \big) \bigg], 
\label{tlagran}
\end{equation} 
where for definitiveness all fields and transformation parameters will be ${\cal
U}(su(2),{\bf ad})$-valued fields in the adjoint representation ${\bf ad}$ of $su(2)$. Here
${\cal U}(su(2),{\bf ad})$ is the universal enveloping algebra of the Lie algebra $su(2)$ of
$SU(2)$. The field $A_{\mu}(x)$ is a Yang-Mills gauge field with field strength
${F}_{\alpha \beta} =\partial _{\alpha}{A}_{\beta}- \partial_{\beta}{A}_{\alpha} +
[{A}_{\alpha}\stackrel{\ast}{,}{A}_{\beta}]$ and $\widetilde{F}^{\alpha \beta} = {1\over 2}
\varepsilon_{\alpha \beta
\gamma \delta} F_{\gamma \delta}$.  The commutator is defined as follows:
$\left[A\stackrel{\ast}{,}B\right]\equiv A\ast B-B\ast A$, and it satisfies the Leibnitz rule
when acting on products of noncommutative fields. The $*$ product is the usual Moyal product
$f(x)\star g(x)\equiv \exp \big(\frac{i}{2}\theta^{\mu\nu}{\frac{\partial }{\partial
y^\alpha}}{\frac{\partial }{\partial y^{\beta}}} \big)  f(y)g(z) \big|_{y=z=x}$ together with
the matrix multiplication. The field contents in $L$ are: the gauge field $A_{\alpha}$, the
ghost fields $\psi_{\alpha}$ and $\phi$, the anti-ghost fields  $\lambda$, $\eta$,
$\chi_{\alpha \beta}$ and an auxiliary field $ H_{\alpha \beta}$. All these fields $(A_{\alpha}
, \psi_{\alpha}, \phi, \lambda, \eta, \chi_{\alpha \beta}, H_{\alpha \beta})$ are ${\cal
U}(su(2),{\bf ad})$-valued fields. Lagrangian (\ref{tlagran}) has a global symmetry identified
with the ghost number $U$. The assignation of $U$ for all these field is respectively:
$(0,1,2,-2,-1,-1,0).$ In the Lagrangian $L$ the covariant derivative $D_{\alpha}$ is defined by
$D_{\alpha} {\bf \Theta} = \partial_{\alpha} {\bf \Theta} + \left[A_{\alpha}\stackrel{\ast}{,}
{\bf \Theta} \right]$.

The Yang-Mills field can be expanded in a basis $\{t^I\}$ of ${\cal U}(su(2),{\bf ad})$ as
$A_{\alpha}(x) = A^I_{\alpha}(x) t^I$ and therefore

\begin{equation}
\left[A_{\alpha}\stackrel{\ast}{,}{A}_{\beta}\right]=
\frac{1}{2}\left\{A^I_{\alpha}\stackrel{\ast}{,}{A}^J_{\beta}\right\}\left[
t_{I},t_{J}\right] +
\frac{1}{2}\left[ A^I_{\alpha}\stackrel{\ast}{,}{A}_{\beta}^{J}\right]
\left\{ t_{I},t_{J}\right\},
\end{equation}
where $\left\{A\stackrel{\ast}{,}B\right\}\equiv A\ast + B\ast A$. Here indices $I,J$ runs over
all possible number of independent generators of
enveloping algebra ${\cal U}(su(2),{\bf ad})$.

Then all the products of the generators $t_I$ will be needed in order to close the algebra
${\cal U}(su(2),{\bf ad})$.
Its structure can be obtained by successively computing the commutators and anticommutators
$\left[ t^{I},t^{J}\right]=f^{IJK}t^{K}$ and $\left\{ t^{I},t^{J}\right\}
= d^{IJK}t^{K}.$

Thus under the assumption $h^{IJK}t^I = t^J \cdot t^K$, the field strength ${F}_{\alpha \beta}$
can be written as

$$
F^I_{\alpha \beta} = \partial _{\alpha}{A}^I_{\beta}- \partial_{\beta}{A}^I_{\alpha} +
h^{IJK} A^J_{\alpha} * {A}^K_{\beta} -  h^{IJK}{A}^K_{\beta} *
A^J_{\alpha} 
$$

\begin{equation}
= \partial _{\alpha}{A}^I_{\beta}- \partial_{\beta}{A}^I_{\alpha} + \frac{1}{2} f^{IJK}
\left\{
A^J_{\alpha}\stackrel{\ast}{,}{A}_{\beta}^{K}\right\}
+ \frac{1}{2} d^{IJK}
\left[
A^J_{\alpha}\stackrel{\ast}{,}{A}_{\beta}^{K}\right],
\label{campoF}
\end{equation}
where $h^{IJK} = {1\over 2} \big(f^{IJK} + d^{IJK}\big)$.

The Lagrangian (\ref{tlagran}) is invariant under a noncommutative BRST-like transformation
\cite{topoym}. Thus the generic noncommutative functionals ${\cal O}$ satisfying the relation
$\delta_{\varepsilon} {\cal O} = -i \varepsilon(x) * \{Q\stackrel{\ast}{,} {\cal O} \}$, where
and $Q$ is the BRST-like charge which it can be easily showed, with the help of \cite{ncbrst},
that it satisfies $Q^2=0$ up to gauge transformations. In general ${\cal O}$ is a functional of
the fields of the theory.

In Refs. \cite{bs,bms} it was showed that fixing the topological symmetry given by the shift
$A_{\alpha} \to A_{\alpha} + \varepsilon_{\alpha} + D_{\alpha} \varepsilon$, there exists a
genuine BRST symmetry which appears after BRST quantize the classical action $\int_X {\rm Tr}
{F}_{\alpha \beta}\widetilde{F}^{\alpha \beta}$. In the procedure in order to enforce the gauge
constraint $\partial \cdot A=0$ it is necessary to introduce a Yang-Mills triplet $(c,
\bar{c},b)$ with corresponding ghost number $U$ to be $(1,-1,0)$. In this noncommutative
case the full BRST symmetry is described by the transformations:

\begin{equation}
\delta A^I_{\alpha}  = D^{\bf ad}_{\alpha}c^I +h^{IJK} \left[ A^J_{\alpha}
\stackrel{\ast}{,} c^K\right], \ \ \ \ \ \ \ 
\delta \psi^I_{\alpha} =
- D^{\bf ad}_{\alpha} \phi^I - h^{IJK} \left[A^J_{\alpha} \stackrel{\ast}{,}
\phi^K\right] - h^{IJK} \left[c^J \stackrel{\ast}{,}
\psi^K_{\alpha} \right],  
\label{brstuno}
\end{equation}
\begin{equation}
\delta \phi^I= -  h^{IJK} \left[c^J, \stackrel{\ast}{,}
\phi^K\right], \ \ \ \ \ 
\delta \chi_{\alpha \beta} = H_{\alpha \beta}, \ \ \ \ \ \  \delta B_{\alpha \beta} =0, \
\ \ \ \ \ \delta \lambda = \eta, \ \ \ \ \ \ \delta \eta =0,
\label{brstdos}
\end{equation}
\begin{equation}
\delta c^I = \phi + {1 \over 2}h^{IJK}\left[c^J\stackrel{\ast}{,} c^K \right], \ \ \ \ 
\delta \bar{c} = b, \ \ \ \ \ \ \delta b =0, 
\label{brsttres}
\end{equation}
where $D^{\bf ad}$ is the
usual covariant derivative in the ${\bf ad}$ representation of the gauge group given by
$D_{\alpha}^{{\bf ad}{IJ}} {\bf \Theta} = \delta^{IJ} \partial_{\alpha} + f^{IJK} A^K_{\alpha}
{\bf \Theta}$. It can be easily showed, following a similar procedure than \cite{ncbrst}
that the BRST charge $Q$ is nilpotent {\it i.e.} $Q^2=0$.

Action (\ref{tlagran}) is BRST-invariant and anomaly free ($\Delta U=0$). If gravity is not
dynamical and only enters as an spectator field and its is coupled to matter only in the usual
commutative way then action can be rewritten as a exact-BRST action {\it i.e.} as a BRST
commutator plus a topological term

\begin{equation}
L'= -{1 \over 4} \int d^4x \sqrt{g} {\rm Tr} F_{\alpha \beta} \widetilde{F}^{\alpha \beta} +  {1
\over e^2} \int \{Q \stackrel{\ast}{,} V\},
\label{siete}
\end{equation}
where
\begin{equation}
V= \int_X d^4 \sqrt{g} {\rm Tr} \bigg[ \lambda * \big(D_{\alpha}\psi^{\alpha} - {1 \over 2}
\beta \eta \big) -  \chi_{\alpha
\beta} * \big( {1\over 2} \alpha H^{\alpha \beta} -{1\over 2} (F^{\alpha \beta} -
\widetilde{F}^{\alpha
\beta})\big) + \bar{c} * \big( \partial \cdot A -{1 \over 2} \gamma b \big)\bigg].
\end{equation}

Lagrangian
(\ref{tlagran}) can be obtained by taking $\alpha =1$ and $\beta = \gamma =0$.  Substituting $V$
into
Eq. (\ref{siete}) and integrating out the auxiliary field $H_{\alpha \beta}$ we can recover 
Lagrangian (\ref{tlagran}). 

In the standard commutative case we have that the energy-momentum tensor $T_{\alpha
\beta}$ of  matter fields is conserved with the usual noncommutative covariant
derivative $D_{\alpha} T^{\alpha \beta} =0$. This energy-momentum tensor is determined 
through its BRST algebra
\cite{topoym} and it is has the form of a BRST commutator. As far as we are considering a
noncommutative deformation of the BRST algebra the corresponding noncommutative version will
have the form of a deformed BRST commutator in the form

\begin{equation}
T_{\alpha \beta} = \{Q \stackrel{\ast}{,} \lambda_{\alpha \beta}\},
\end{equation}
where
$$
\lambda_{\alpha \beta} = {1\over 2} {\rm Tr} \bigg[ F_{\alpha \sigma} * \chi_{\beta}^{ \
\sigma}  +  F_{\beta \sigma} * \chi_{\alpha}^{ \ \sigma} - {1\over 2} g_{\alpha \beta} F_{\mu
\nu} * \chi^{\mu \nu}\bigg] 
$$
\begin{equation}
+ {1\over 2}{\rm Tr}  \bigg( \psi_{\alpha} * D_{\beta} \lambda +
\psi_{\beta}*  D_{\alpha} \lambda - g_{\alpha \beta} \psi_{\sigma} * D^{\sigma}\lambda \bigg) +
{1\over 4} g_{\alpha \beta} {\rm Tr} \big(\eta * \big[ \phi \stackrel{\ast}{,} \lambda \big]
\big).
\end{equation}
Of course the energy-momentum tensor of the noncommutative matter is conserved with the
noncommutative covariant derivative $D_{\alpha} {\bf \Theta} = \partial_{\alpha} {\bf \Theta} +
\left[A_{\alpha}\stackrel{\ast}{,}
{\bf \Theta} \right]$.

\section{Observables}

Physical observables of the topological theory are BRST cohomology classes of $Q$. From the
previous section
we have that the scalar field $\phi^I(x)$ is a BRST invariant since $\left[Q\stackrel{\ast}{,}
\phi^I(x) \right] =0$. Any gauge invariant polynomial of this field is a physical observable,
for instance take

\begin{equation}
{\cal O}_{k,0}(x) = {\rm Tr} \phi_*^k \equiv {\rm Tr} \big( \phi(x) *  \phi(x) * \dots *
\phi(x)
\big), \ \ \  k \
{\rm times}
\end{equation}
with ghost number $U=2k$. 

Using the properties of $X_{\theta}$, the property of the differential operator on $X_{\theta}$
{\it i.e.} $d (f_1 * f_2) = df_1 * f_2 + f_1 * df_2$ (with $f_1,f_2 \in X_{\theta}$) and the BRST
algebra
(4) we get

\begin{equation}
{\partial {\cal O}_{k,0} \over \partial x^{\alpha}}  = k {\rm Tr} \phi_*^{k-1} * D_{\alpha}
\phi(x) = \{Q \stackrel{\ast}{,} -k {\rm Tr} \phi_*^{k-1} * \psi_{\alpha}(x) \}.
\label{observa}
\end{equation}
We will use the standard definition of noncommutative wedge product $\stackrel{\ast}{\wedge}$

\begin{equation}
{\bf F} \stackrel{\ast}{\wedge} {\bf G} \equiv F_{\alpha_1 \dots \alpha_p}(x) * G_{\beta_1 \dots
\beta_q} dx^{\alpha_1} \wedge \dots \wedge  dx^{\alpha_p}\wedge
dx^{\beta_1} \wedge \dots \wedge dx^{\beta_q}.
\end{equation}
where ${\bf F} \in \Lambda^p(T^* X_{\theta})$ and ${\bf G} \in \Lambda^q(T^* X_{\theta})$.

We find that we can rewrite the Eq. (\ref{observa}) as 

\begin{equation}
d{\cal O}_{k,0} = \{Q \stackrel{\ast}{,} {\cal O}_{k,1} \},
\end{equation}
where $  {\cal O}_{k,1}= -k {\rm Tr} \phi_*^{k-1} * \psi_{\alpha}(x) dx^{\alpha}.$ In this
noncommutative extension of the BRST symmetry the observables ${\cal O}_{k,a}$ with
$a=0,1,2,3,4$ also generate noncommutative BRST-invariant operators $W_k({\cal S}_a)$ where
${\cal S}_a$ is a homology cycle of $X$ of dimension $a$. These operators are defined as

\begin{equation}
W_k({\cal S}_a) = \int_{{\cal S}_a} {\cal O}_{k,a}.
\end{equation}
It is easy to show that in the noncommutative BRST algebra we have

\begin{equation}
\{ Q \stackrel{\ast}{,} W_k({\cal S}_a) \} = \int_{{\cal S}_a} \{ Q \stackrel{\ast}{,} {\cal
O}_{k,a} \} = \int_{{\cal S}_a} d {\cal O}_{k,a} = 0.
\end{equation}
Thus  $W_k({\cal S}_a)$ are BRST-invariant operators and they are organized in BRST cohomology
classes of the BRST charge $Q$. 

The rest of the operators ${\cal O}_{k,a}$ with $a>1$ can be computed easily. Thus we
have 

\begin{equation}
{\cal O}_{k,2} = {\rm Tr} \big( i \phi \stackrel{\ast}{\wedge} F + {1\over 2} \psi
\stackrel{\ast}{\wedge} \psi \big),
\end{equation}
\begin{equation}
{\cal O}_{k,3} = i {\rm Tr} \big(\psi \stackrel{\ast}{\wedge} F \big),
\end{equation}
\begin{equation}
{\cal O}_{k,4} = -{1\over 2} {\rm Tr} \big( F \stackrel{\ast}{\wedge} F \big).
\end{equation}

In the commutative case $[{\cal O}_{k,a}]$ are organized in cohomology classes of de Rham
cohomology.  $W_k({\cal S}_a)$ gives a pairing relating ${\cal O}_{k,a}$ to homology cycles
$[{\cal S}_a]$ by Poincar\'e duality. In the noncommutative case  $[{\cal O}_{k,a}]$ defines
cohomology classes in the Chevalley-Eilenberg cohomology $H^{\bullet}(X_{\theta})$ of the Moyal
algebra
$X_{\theta}$. By Poincar\'e duality classes $[{\cal O}_{k,a}]$ can be coupled to
Chevalley-Eilenberg homology cycles $[{\cal S}_a] \in H_{\bullet}(X_{\theta})$.

\section{Noncommutative Donaldson Invariants}

\subsection{Moduli Space of Noncommutative Instantons}

As in the usual commutative case, in the noncommutative theory the partition function $Z$ is
given by

\begin{equation}
Z= \int ({\cal D}X) \exp \big( - L'/e^2\big),
\end{equation}
where the measure ${\cal D}X$ is a notation to abbreviate ${\cal D} A^I_{\alpha} \cdot
{\cal
D} \psi^I_{\alpha} \cdot {\cal D} \phi^I \cdot \dots \cdot {\cal D}\chi$.

In the case when the gravity is an spectator and it is not coupled noncommutatively to
the matter fields, partition function $Z$ 
is a topological invariant of the noncommutative
manifold $X_{\theta}$. To see that we can use the fact that $L'$ is a
$\theta$-deformed BRST commutator (see Eq. (\ref{siete})). These invariants are independent on
the
metric but they
may depend on the differentiable structure of the noncommutative manifold
$X_{\theta}$. In this section we will argued that these topological invariants would be
considered as the noncommutative deformation of
Donaldson polynomial invariants of $X_{\theta}$. 

For the moduli space $\widehat{\cal M}$ of zero dimension, the only topological invariant is
the
partition function $Z$. For positive dimension the non-vanishing
topological invariants are the correlation functions given by

\begin{equation}
Z({\cal O}) = \langle {\cal O} \rangle = \int ({\cal D}X) \exp \big( - L'/e^2\big) \cdot {\cal
O},
\end{equation}
where ${\cal O}$ is a functional of the involved noncommutative fields. 

The statement of that $\langle \{Q,{\cal O} \} \rangle =0$ (for any ${\cal O}$) valid in the
commutative theory is still valid in the noncommutative case since the necessary and sufficient
condition is that our noncommutative Lagrangian (\ref{siete}) be a noncommutative BRST invariant
and therefore $\{Q \stackrel{\ast}{,} L' \} = 0$ and therefore $\langle \{Q
\stackrel{\ast}{,}{\cal O} \} \rangle =0$. For similar reasons the correlation function $\langle
A * \{Q \stackrel{\ast}{,} B \} \rangle =0$ for any functional $B$ if $ \{Q \stackrel{\ast}{,} A
\} =0$ for any functional $A$.

Partition function and correlation functions are also independent of the gauge coupling
constant
$e$. Hence we can evaluate the path integral for small values of $e^2$ {\it i.e.} $e^2 \to 0$.
The application of the stationary phase approximation leads to the fact that  
the path integral is dominated by the classical minima and thus it is concentrated
precisely at the noncommutative instanton field configuration \cite{ns,schwarz}

\begin{equation}
{F}^I_{\alpha \beta} = - \widetilde{F}^I_{\alpha \beta}.
\end{equation}
Solutions to this equation define the moduli space of noncommutative instantons $\widehat{\cal
M}^{\theta}_{inst}$ which is a smooth manifold already without small instanton singularities
\cite{ns,schwarz,sw}. In this case the self-duality condition is deformed by the last term of
the Eq. (\ref{campoF}) given by $ \frac{1}{2} d^{IJK} \left[
A^J_{\alpha}\stackrel{\ast}{,}{A}_{\beta}^{K}\right]$. As parameter $\theta \to 0$ this term
goes also to zero. This is precisely the term that deforms the ADHM equations of
\cite{ns,schwarz}.

\subsection{Noncommutative Deformation of Moduli Space}

In the commutative case the moduli space of instantons ${\cal M}_{inst}$ has singularities where
the size of instantons goes to zero, they are called small instanton singularities which are 
related to the existence of zero modes for ghost fields. Then when
one integrate out on this space it is usually assumed that they are no present. However in the
noncommutative case these singularities are absent because of the existence of a length scale
provided by the noncommutative
parameter $\theta$ \cite{ns,schwarz}. Thus it is not necessary of dropping out the zero modes of
other fields than gauge field $A^I_{\alpha}$ and $\psi^I_{\alpha}$.  After evaluating 
Feynman integral in the weak coupling limit ($e \to 0$) and integrating out the
non-zero bosonic and fermionic modes  we get from the
stationary phase approximation that

\begin{equation}
Z({\cal O}) = \int_{\widehat{\cal M}^{\theta}_{inst}} d a_1 \dots da_n d\psi_1 \dots d\psi_n \
{\cal O},
\end{equation}
where $(a_1,a_2, \dots a_n)$ are the coordinates of the moduli space of
noncommutative instantons $\widehat{\cal M}^{\theta}_{inst}$ and  ${\cal
O}$ being any BRST-invariant functional. The part of ${\cal O}$ depending only on
its zero modes is given by

\begin{equation}
{\cal O}' = \Phi_{i_1 \dots i_n}(a^k) \cdot \psi^{i_1} *  \psi^{i_2} \dots * \psi^{i_n}.
\end{equation}
Then correlation function of the functional ${\cal O}$ leads to an integration of the a
differentiable closed $n$-form $\widehat{\Phi}$ on the moduli space $\widehat{\cal
M}^{\theta}_{inst}$

$$
Z({\cal O}) = \int_{\widehat{\cal M}^{\theta}_{inst}} d a_1 \dots da_n d\psi_1 \dots d\psi_n
\Phi_{i_1 \dots
i_n}(a^k) \cdot
\psi^{i_1} * \dots * \psi^{i_n}
$$
\begin{equation}
= \int_{\widehat{\cal M}^{\theta}_{inst}} \widehat{\Phi}.
\end{equation}

If we take the Moyal product of $k$ functionals of fields we have 
${\cal O} = {\cal O}_1 * {\cal O}_2*  \dots * {\cal O}_k,$ where each ${\cal O}_r$ has
$U=n_r$ such that $n =\sum_r^k n_r$. After integrating out the zero modes
we get

\begin{equation}
{\cal O}'_r = {\Phi}^{(r)}_{i_1 \dots i_{n_r}} \psi^{i_1} * \dots * \psi^{i_{n_r}}.
\end{equation}

This leads to reexpress the product as ${\cal O}' = {\cal O}'_1 * {\cal
O}'_2* \dots * {\cal O}'_k$. Then correlation functions are given by

\begin{equation}
Z({\cal O}_{{\alpha}_1}* {\cal O}_{{\alpha}_2}* \dots *{\cal O}_{{\alpha}_s}) 
= \int_{\widehat{\cal M}}({\cal D} X) \exp \big( -L'/e^2\big) {\cal
O}_{{\alpha}_1} * {\cal O}_{{\alpha}_2} * \dots * {\cal O}_{{\alpha}_s}.
\end{equation}
This leads to express the topological invariants, in the limit of $e \to 0$, as an integral in
the moduli space of
noncommutative instantons of the wedge product of differentiable forms 

\begin{equation}
Z({\cal O}_{{\alpha}_1}* {\cal O}_{{\alpha}_2}* \dots * {\cal
O}_{{\alpha}_s}) = \int_{\widehat{\cal M}^{\theta}_{inst}} \widehat{\Phi}^{(\alpha_1)}
\buildrel{*}\over{\wedge}  \widehat{\Phi}^{(\alpha_2)} \buildrel{*}\over{\wedge}
\dots \buildrel{*}\over{\wedge} \widehat{\Phi}^{(\alpha_n)},
\label{ncdi}
\end{equation}
where now the wedge product $\stackrel{\ast}{\wedge}$ is defined on $\widehat{\cal
M}^{\theta}_{inst}$.

From the previous section we described that for Donaldson theory the relevant functionals were
of the form ${\cal O}^{({\cal S}_a)} \equiv W_k({\cal S}_a) =\int_{{\cal S}_a} {\cal O}_{k,a}$,
where $[{\cal S}_a]$ is the Chevalley-Eilenberg homology cycle of dimension $a$ in the algebra
$X_{\theta}$. Functional $ {\cal O}_{k,a}$ has ghost number $4-a$ and thus the functional ${\cal
O}^{({\cal S}_a)}$ is associated with a $4-a$ form $\widehat{\Phi}^{(4-a)}$ in the moduli space
$\widehat{\cal M}^{\theta}_{inst}$. Then we have for these functionals that

\begin{equation}
Z({\cal O}^{({\cal S}_1)}* {\cal O}^{({\cal S}_2)} * \dots * {\cal O}^{({\cal S}_{\ell})}) =
\int_{\widehat{\cal M}^{\theta}_{inst}} \widehat{\Phi}^{({\cal S}_1)}
\buildrel{*}\over{\wedge}  \widehat{\Phi}^{({\cal S}_2)} \buildrel{*}\over{\wedge}
\dots \buildrel{*}\over{\wedge} \widehat{\Phi}^{({\cal S}_{\ell})}.
\label{ncdidos}
\end{equation}
This topological invariant can be regarded as 

\begin{equation}
Z({\cal S}_1, {\cal S}_2, \dots  ,{\cal S}_{\ell}) =
\int_{\widehat{\cal M}^{\theta}_{inst}} \widehat{\Phi}^{({\cal S}_1)}
\buildrel{*}\over{\wedge}  \widehat{\Phi}^{({\cal S}_2)} \buildrel{*}\over{\wedge}
\dots \buildrel{*}\over{\wedge} \widehat{\Phi}^{({\cal S}_{\ell})}.
\label{ncditres}
\end{equation}

Equation (\ref{ncditres}) provides a map from Chevalley-Eilenberg homology cycles $[{\cal S}_a]$ 
of $X_{\theta}$ to the Chevalley-Eilenberg
cohomology cycles of the moduli space seen as a noncommutative smooth space $\widehat{\cal
M}^{\theta}_{inst}$. In this case we have that the
Donaldson map is given by
$Z: H_a(X_{\theta}) \to H^{4-a}(\widehat{\cal M}^{\theta}_{inst})$ and they represent a
noncommutative
deformation of Donaldson invariants given by integral (\ref{ncditres}), which can be interpreted
as the intersection form of $\widehat{\cal M}^{\theta}_{inst}.$

\section{Concluding Remarks}

In this paper we have proposed a noncommutative deformation of Witten's cohomological field
theory for Yang-Mills gauge theory \cite{topoym}. This deformation is performed by deforming
the underlying Lie algebra and changing it by the enveloping algebra following \cite{wess}. The
application of these ideas leads us to a consistent noncommutative deformation of the full
BRST fermionic symmetry of this topological theory given by Eqs. (\ref{brstuno}), (\ref{brstdos})
and (\ref{brsttres}). The results of
noncommutative deformations of a standard BRST quantization of Ref. \cite{ncbrst} help us to
show that in our case BRST charge is nilpotent i.e. $Q^2 =0$. 

Observables of this deformed theory are computed and showed to be BRST and gauge invariant in the
noncommutative sense. Partition function and correlation functions of these operators are proved
to be topological invariants and independent of the gauge coupling $e$ as in the standard
commutative case. Weak coupling limit ($e \to 0$) and the stationary phase approximation are used
to show that the main contribution of Feynman integrals gives precisely the condition that
satisfies the noncommutative Yang-Mills instantons \cite{ns,schwarz}.

Gathering all that we are able to define deformed Donaldson invariants as integrals on the
noncommutative moduli space of noncommutative differential forms. This form can be interpreted as
the deformed Donaldson map $Z: H_a(X_{\theta}) \to H^{4-a}(\widehat{\cal M}^{\theta}_{inst})$,
given by a mapping between the Chevalley-Eilenberg homology of noncommutative spacetime
$X_{\theta}$ and the Chevalley-Eilenberg cohomology of the noncommutative moduli space of
instantons $\widehat{\cal M}^{\theta}_{inst}$.


\vskip 2truecm
\centerline{\bf Acknowledgments}
This work was supported in part by CONACyT M\'exico Grant
33951E. P.P. is supported by a CONACyT graduate fellowship. It is a pleasure to thank A.
Dymarsky, O. Obreg\'on and C. Ram\'{\i}rez for very useful discussions.


\vskip 2truecm



\end{document}